\RequirePackage{fix-cm}
\documentclass[twocolumn,epjc3]{svjour3}
\smartqed  % flush right qed marks, e.g. at end of proof
\usepackage{amsmath}
\usepackage{amssymb}
\usepackage{amssymb,bbold}

\RequirePackage{graphicx}
\RequirePackage[colorlinks,citecolor=blue,urlcolor=blue,linkcolor=blue]{hyperref}
\RequirePackage[numbers,sort&compress]{natbib}

\journalname{Eur. Phys. J. C}
\begin{document}

\title{Comment on ``Effects of cosmic-string framework on the thermodynamical properties of anharmonic oscillator using the ordinary statistics and the $q$-deformed superstatistics approaches''%\thanksref{t1}
}


\author{Francisco A. Cruz Neto\thanksref{e1,addr1} \and Luis B. Castro\thanksref{e2,addr1} %etc.
}

\thankstext{e1}{e-mail: alvesfcn@gmail.com}
\thankstext{e2}{e-mail: luis.castro@ufma.br, luis.castro@pq.cnpq.br}


\institute{Departamento de F\'{\i}sica, Universidade Federal do Maranh\~{a}o, Campus Universit\'{a}rio do Bacanga, 65080-805, S\~{a}o Lu\'{\i}s, MA, Brazil.
\label{addr1}
}

\date{Received: date / Accepted: date}

\maketitle

\begin{abstract}
We point out a misleading treatment in a recent paper published in this Journal [Eur. Phys. J. C (2018)78:106] regarding solutions for the Schr\"{o}dinger equation with a anharmonic oscillator potential embedded in the background of a cosmic string mapped into biconfluent Heun equation. This fact jeopardizes the thermodynamical properties calculated in this system.

%\PACS{04.62.+v \and 04.20.Jb \and 03.65.Pm \and 03.65.Ge}
\end{abstract}

%\section{Introduction}
%\label{intro}

In a recent paper in this Journal, Sobhami et. al. \cite{EPJC78:106:2018} have studied the thermodynamical properties of the anharmonic oscillator within cosmic-string framework using ordinary statistic and the $q$-deformed superstatistics approaches. To achieve their goal, the authors need to calculate the wave function and the energy spectrum, which have been obtained from the Schr\"{o}dinger equation within a cosmic-string framework mapped into biconfluent Heun differential equation. It is worthwhile to mention that all results depend mainly on the energy spectrum of the system. The purpose of this comment is point to out a misleading treatment on the solution of the biconfluent Heun equation, this fact jeopardizes the results of \cite{EPJC78:106:2018}. 

The time-independent Schr\"{o}dinger equation with an anharmonic oscillator potential embedded in the background of a cosmic string is given by
\begin{equation}\label{eq1}
\begin{split}
\frac{d^2\Phi(\rho)}{d\rho^2}+\frac{1}{\rho}\frac{d\Phi(\rho)}{d\rho}+\left( \varepsilon-k_{z}^{2}-\left( a_{v}+l^{2}\alpha^{2} \right)\rho^2 \right. \\ \left.-b_{v}\rho^{4}-c_{v}\rho^{6} \right)\Phi(\rho)=0\,.
\end{split}
\end{equation}
\noindent Redefining the wave function as $\Phi(\rho)=\frac{R(\rho)}{\sqrt{\rho}}$, one can remove the first derivative and rewrite Eq.(\ref{eq1}) as
\begin{equation}\label{eq2}
\begin{split}
\frac{d^{2}R(\rho)}{d\rho^{2}}+\left( \varepsilon+\frac{1}{4\rho^{2}}-k_{z}^{2}-\left( a_{v}+l^{2}\alpha^{2} \right)\rho^2 \right. \\
\left. -b_{v}\rho^{4}-c_{v}\rho^{6} \right)R(\rho)=0\,.
\end{split}
\end{equation} 
\noindent Making use of the new variable $y=\rho^{2}$ and redefining the wave function as $R(y)=\frac{f(y)}{\sqrt[4]{y}}$, the Eq.(\ref{eq2}) becomes
\begin{equation}\label{etsch}
\frac{d^{2}f(y)}{dy^{2}}+\left( \kappa+\frac{A}{y}-By-Cy^{2}+\frac{1/4}{y^{2}} \right)f(y)=0\,,
\end{equation}
\noindent where
\begin{eqnarray}
\kappa &=& -\frac{1}{4}\left( a_{v}+l^{2}\alpha^{2} \right)\,,\\
A &=& \frac{1}{4}\left( \varepsilon-k_{z}^{2} \right)\,,\label{pa_A}\\
B &=& \frac{b_{v}}{4}\,,\\
C &=& \frac{c_{v}}{4}\,.
\end{eqnarray}
\noindent The solution for the differential equation (\ref{etsch}) with $C$ necessarily real and positive ($c_{v}>0$), is the solution of the Schr\"{o}\-dinger equation for the three-dimensional harmonic oscillator plus a Cornell potential \cite{JPA19:3527:1986,RONVEAUX1995,PRC86:052201:2012}. It is worthwhile to mention that Refs.\cite{JPA19:3527:1986,RONVEAUX1995} present some erroneous calculations.

Considering the solution in the form of \cite{RONVEAUX1995}
\begin{equation}\label{solh}
f(y)=y^{1/2}\exp\left( -\frac{\sqrt{c_{v}}}{4}y^{2}-\frac{b_{v}}{4\sqrt{c_{v}}}y \right)F(y)\,,
\end{equation}  
\noindent and by introducing the new variable and parameters:
\begin{eqnarray}
x &=& \left(\frac{c_{v}}{4}\right)^{1/4}y\,,\label{newv}\\
\beta^{\prime} &=& \left(\frac{4}{c_{v}}\right)^{3/4}\frac{b_{v}}{4}\,,\\
\gamma^{\prime} &=& \left(\frac{4}{c_{v}}\right)^{3/2}\left[ \frac{b_{v}^{2}}{64}-\frac{c_{v}}{16}\left( a_{v}+l^{2}\alpha^{2} \right) \right]\,,
\end{eqnarray}
\noindent one finds that the solution can be expressed as a solution of the biconfluent Heun differential equation \cite{EPJC72:2051:2012,PLA376:2838:2012,AP341:86:2014,AP355:48:2015,EPJC78:44:2018}
\begin{equation}
\begin{split}
\frac{d^{2}F(x)}{dx^{2}}+\left( \frac{1}{x}-\beta^{\prime}-2x \right)\frac{dF(x)}{dx} \\
+\left[ \gamma^{\prime}-2-\frac{\Theta}{x} \right]F(x)=0\,,
\end{split}
\end{equation}
\noindent with
\begin{equation}
\Theta=\frac{1}{2}\left[ \delta^{\prime}+\beta^{\prime} \right]\,,
\end{equation}
\noindent where
\begin{equation}
\delta^{\prime}=\left(\frac{4}{c_{v}}\right)^{1/4}\frac{k_{z}^{2}-\varepsilon}{2}\,.
\end{equation}
\noindent This differential equation has a re\-gular singularity at $x=0$ and an irregular singularity at $x=\infty$. The regular solution at the origin is given by
\begin{equation}
H_{b}\left(0,\beta^{\prime},\gamma^{\prime},\delta^{\prime};x\right)=\sum_{j=0}^{\infty}\frac{1}{\Gamma(1+j)}
\frac{A_{j}}{j!}x^{j}\,,
\end{equation}
\noindent here $\Gamma(z)$ is the gamma function, $A_{0}=1$, $A_{1}=\Theta$ and the remaining coefficients for $\beta^{\prime}\neq 0$ satisfy the recurrence relation,
\begin{equation}
\begin{split}
A_{j+2}=\left[ (j+1)\beta^{\prime}+\Theta \right]A_{j+1}-(j+1)^{2}(\Delta-2j)A_{j}\,,
\end{split}
\end{equation}
\noindent where $\Delta=\gamma^{\prime}-2$.

The series is convergent and tends to $\exp\left(x^2+\beta^{\prime}x\right)$ as $x\rightarrow\infty$. It is true that the presence of $\exp\left(x^2+\beta^{\prime}x\right)$ in the asymptotic behavior of $H_{b}\left(0,\beta^{\prime},\gamma^{\prime},\delta^{\prime};x\right)$ perverts the normalizability of the solution $f(y)$, i.e $f(y)\propto \exp\left(\frac{\sqrt{c_{v}}}{4}y^{2}+\frac{b_{v}}{4\sqrt{c_{v}}}y \right)$ as $y\rightarrow\infty$. Nevertheless, this trouble can be surpassed by considering a polynomial solution for $H_{b}\left(0,\beta^{\prime},\gamma^{\prime},\delta^{\prime};x\right)$. In fact, $H_{b}\left(0,\beta^{\prime},\gamma^{\prime},\delta^{\prime};x\right)$ presents polynomial solutions of degree $n$ if and only if two conditions are satisfied:
\begin{equation}\label{c1}
\Delta=2n, \qquad n=0,1,2,\ldots
\end{equation}
\noindent and
\begin{equation}\label{c2}
A_{n+1}=0\,.
\end{equation}
\noindent Now, the condition (\ref{c2}) provides a polynomial of degree $n+1$ in $\delta^{\prime}$ and there are at most $n+1$ suitable values of $\delta^{\prime}$. Therefore, the energy eigenvalues of the system are obtained for the both conditions (\ref{c1}) and (\ref{c2}).

From the condition (\ref{c1}), one obtains
\begin{equation}\label{bn}
b_{v,n}=\pm\left(\frac{c_{v}}{4}\right)^{3/4}\sqrt{128(n+1)+\sigma}\,,
\end{equation}
\noindent where 
\begin{equation}
\sigma=16\left(\frac{4}{c_{v}}\right)^{1/2}\left(a_{v}+l^2\alpha^2\right)\,.
\end{equation}
\noindent This result shows a constraint that involves a specific value for $b_{v}$ as a function of $n$, $c_{v}$, $a_{v}$, $l$ and $\alpha$. The problem does not end here, it is necessary to analyze the second condition of quantization.

Now, we focus attention on the condition (\ref{c2}). For $n=0$, the condition $A_{1}=\Theta=0$ results in an algebraic equation of degree one in $\delta^{\prime}$: 
\begin{equation}
\delta^{\prime}+\beta^{\prime}=0\,,\nonumber
\end{equation}
\noindent which furnishes the following expression for the energy
\begin{equation}\label{e0}
\varepsilon_{0}=k_{z}^{2}+ \frac{b_{v,0}}{2}\left(\frac{4}{c_{v}}\right)^{1/2}\,.
\end{equation}
\noindent Note that this result presents two different classes of solutions that depend on the sign of $b_{v,0}$. If $A\gtrless 0$ [Eq.(\ref{pa_A})], that implies $\varepsilon\gtrless k_{z}^{2}$, then $b_{v,0}\gtrless 0$, so we get 
\begin{equation}\label{e0_pm}
\varepsilon_{0,\pm}=k_{z}^{2}\pm\frac{1}{2}\left(\frac{c_{v}}{4}\right)^{1/4}\sqrt{128+\sigma}\,.
\end{equation}
\noindent The expression (\ref{e0_pm}) represents the energy eigenvalue for $n=0$.

Now, let us consider the case for $n=1$, which implies that $A_{2}=\left(\beta^{\prime}+\Theta\right)\Theta=0$. In this case we obtain an algebraic equation of degree two in $\delta^{\prime}$:
\begin{equation}
\left(\delta^{\prime}+\beta^{\prime}\right)^{2}+2\beta\left(\delta^{\prime}+\beta^{\prime}\right)=0\,, \nonumber
\end{equation}
\noindent which furnishes the energy eigenvalues 
\begin{equation}\label{e1_1}
\varepsilon_{1,1}=k_{z}^{2}+ \frac{b_{v,1}}{2}\left(\frac{4}{c_{v}}\right)^{1/2}\,
\end{equation}
\noindent and
\begin{equation}\label{e1_2}
\varepsilon_{1,2}=k_{z}^{2}+ \frac{3b_{v,1}}{2}\left(\frac{4}{c_{v}}\right)^{1/2}\,.
\end{equation}
\noindent For simplicity, we will only consider the case $b_{v,1}>0$ [positive sign in (\ref{bn})]. Therefore, the energy eigenvalues in this case are given by
\begin{equation}\label{e1_1c}
\varepsilon_{1,1,+}=k_{z}^{2}+\frac{1}{2}\left(\frac{c_{v}}{4}\right)^{1/4}\sqrt{256+\sigma}\,.
\end{equation}
\noindent and
\begin{equation}\label{e1_2c}
\varepsilon_{1,2,+}=k_{z}^{2}+\frac{3}{2}\left(\frac{c_{v}}{4}\right)^{1/4}\sqrt{256+\sigma}\,.
\end{equation}

For the case of $n=2$, the condition $A_{3}=0$ results in an algebraic equation of degree three in $\delta^{\prime}$. And the solutions, in fact, turn out to be three values for the energy (considering $b_{v,3}>0$). For $n\geq 2$ the algebraic equations are cumbersome. With this result, we conclude that the energy can not be labeled with $n$. This is a peculiar behavior of the biconfluent Heun equation. Our results show that the expression for the energy in Ref.\cite{EPJC78:106:2018} is wrong, probably due to erroneous calculations in the manipulation of the biconfluent Heun equation. The correct quantization condition is obtained applying two conditions: the condition (\ref{c1}) is used to obtain constraints between the potential parameters and the condition (\ref{c2}) is used to obtain the expression of energy eigenvalues. It is worthwhile mention that condition of quantization is different for each value of $n$.
 
In summary, we analyzed the solution for the Schr\"{o}\-dinger equation with a anharmonic oscillator potential embedded in the background of a cosmic string. In this process, the problem is mapped into biconfluent Heun differential equation and using appropriately the conditions (\ref{c1}) and (\ref{c2}), we found the correct energy eigenvalues and constraints on the potential parameters. Also, we showed that there is no need for fixing the value of $c_{v}=4$ for obtain a biconfluent Heun equation, in contrast to Ref.\cite{EPJC78:106:2018}. Additionally, the thermodynamical properties calculated in Ref.\cite{EPJC78:106:2018} depend of the energy spectrum relation, therefore our results jeopardize the main results of Ref.\cite{EPJC78:106:2018}.

\begin{acknowledgements}
We acknowledge valuable comments from the anonymous referee. This work was supported in part by means of funds provided by CNPq, Brazil, Grant No. 307932/2017-6 (PQ).
\end{acknowledgements}

\bibliographystyle{spphys}       % APS-like style for physics

%\bibliography{mybibfile_stars2}

\end{document}